\documentclass[a4paper]{article}
\usepackage[natbib=true]{biblatex}
\usepackage{dmbiblatex}
\usepackage{hyperref}
\usepackage{authblk,orcidlink}

\title{Memory Simulations, Security and Optimization in a Verified Compiler}

\newcommand{\compcert}{CompCert}
\newcommand{\gcc}{\textsf{gcc}}
\newcommand{\llvm}{LLVM}
\newcommand{\clang}{\textsf{clang}}
\newcommand{\Vundef}{\textsf{Vundef}}
\newcommand{\chyphen}{\discretionary{}{-}{-}}

\newcommand{\official}[2]{\href{https://github.com/AbsInt/CompCert/commit/#2}{{\small\texttt{#1}}}}
\newcommand{\chamois}[2]{\href{https://gricad-gitlab.univ-grenoble-alpes.fr/certicompil/Chamois-CompCert/-/commit/#2}{{\small\texttt{#1}}}}

\usepackage{listings}
\newcommand{\module}[2]{\href{https://compcert.org/doc/html/compcert.#1.#2.html}{\texttt{#2.v}}}
\newcommand{\smallfigurefontsize}{\fontsize{5}{6}\selectfont}

\newcommand*{\Description}[2][]{}
\newcommand*{\grantsponsor}[3]{\href{#3}{#2}}
\newcommand*{\grantnum}[3][]{\href{#1}{#3}}

\bibliography{memory_simulation}

\begin{document}

\author{David Monniaux\orcidlink{0000-0001-7671-6126}}
\affil{Univ. Grenoble Alpes, CNRS, Grenoble INP\footnote{Institute of Engineering Univ. Grenoble Alpes}, VERIMAG, 38000 Grenoble, France.}

\maketitle

\begin{abstract}
Current compilers implement security features and optimizations that require nontrivial semantic reasoning about pointers and memory allocation: the program after the insertion of the security feature, or after applying the optimization, must simulate the original program despite a different memory layout.
  
In this article, we illustrate such reasoning on pointer allocations through memory extensions and injections, as well as fine points on undefined values, by explaining how we implemented and proved correct two security features (stack canaries and pointer authentication) and one optimization (tail recursion elimination) in the CompCert formally verified compiler.

\end{abstract}

\section{Introduction}
Formally verified compilation aims at providing an end-to-end mathematical, machine-checked proof of preservation of the semantics of code through the compilation process.
There are currently only two formally verified compilers for general-purpose languages: 
{\compcert} \cite{DBLP:journals/cacm/Leroy09}%
\footnote{{\compcert} is distributed both as source code, free of charge for research purposes, to which we will refer when discussing CompCert modules, available at \url{https://github.com/AbsInt/CompCert} and as a commercial package by Absint \url{https://www.absint.com/}}, for almost all of the C programming language,%
\footnote{There also exist other front-ends for {\compcert} for languages other than~C. \href{https://velus.inria.fr/}{V{\'e}lus} is a front-end for a subset of the Lustre or Scade synchronous programming languages. A front-end for a subset of OCaml was also developed \cite{DBLP:phd/hal/Dargaye09}.}
and CakeML for a substantial subset of Standard~ML.%
\footnote{%
In addition, some optimization passes for LLVM were formalized with the Vellvm project~\cite{Zhao_PhD2013,DBLP:conf/popl/ZhaoNMZ12,DBLP:conf/pldi/NagarakatteZMZ09}, but this was not an end-to-end project.}

Our goal in this paper is to illustrate how certain proofs methods for reasoning on semantics and transformations of programs involving pointers can be used to implement optimization and security features not found in these compilers.

In this paper, we shall be concerned with {\compcert}. {\compcert} is used for safety-critical applications, including fly-by-wire controls in commercial aircraft \cite{Leroy-PPES-2011}.
It however does not support many optimizations and security features found in current compilers such as {\gcc} and {\llvm} (\clang), which may limit its appeal for other contexts.
One reason is that implementing and proving these features correct is considered challenging, particularly if \emph{memory} is involved.

The internal representations (IR) of many compilers, including {\llvm} and {\compcert}, distinguish between values found in (pseudo-)registers and in memory. Registers are CPU registers and thus are in limited number, while infinitely many pseudo-registers are available. The \emph{register allocation} phase, occurring after most optimizations have been run, maps pseudo-registers to CPU registers or to stack locations and adds suitable memory load and store instructions to \emph{spill} values to them. Conversely, some early compilation phase (in \llvm, the \textsf{mem2reg} phase) tries to map local variables to pseudo-registers, if possible, otherwise to stack memory locations. The difference is that the address of pseudo-registers cannot be taken, they cannot be referred to through pointers, and there is no possibility of aliasing (two pseudo-registers are identical if and only if they have the same identifier, while memory locations may be identical or not according to complex pointer arithmetic relationships).

In {\compcert}, the correctness of code transformations and optimization phases is established by exhibiting a ``match'' relation between the program states before and after the transformation and proving that sequences of steps of the program pre-transformation are matched by sequences of steps post-transformation, a form of \emph{simulation}. The overall correctness of the compiler is established by composition of these simulations.

In this paper, we will show how to use forms of memory simulations, known as memory extensions and injections \cite{DBLP:conf/esop/BeringerSDA14}, to prove the correctness of certain compiler transformations that are available in mainstream compilers yet missing from official releases of {\compcert}:
\begin{enumerate}
\item
Official releases of {\compcert} do feature \emph{tail call elimination}, which
will in particular turn some recursive calls into tail recursive calls. However these tail recursive calls will still restore and save registers, destroy and reconstruct stack frames. Tail recursion elimination improves on tail call elimination by bypassing the function epilogue and prologue by directly branching to the top of the function body (after, of course, placing correct values in the parameters). The optimization itself is rather trivial to implement, but its correctness proof is nontrivial; it illustrates the use of \emph{memory injections}.
\item
\emph{Stack canaries} are a basic but effective countermeasure to certain buffer overflow attacks and have long been supported by {\gcc } and {\llvm}. Again, the transformation is trivial but the correctness proof involves \emph{memory extensions}.
\item
\emph{Return address authentication} is another countermeasure to certain buffer overflows, supported on recent ARM processors. Here, suitably limited modification to memory invariants enable supporting this feature.
\end{enumerate}

An important aspect, when dealing with formally verified software, is its \emph{trusted computing base} (TCB).
We extensively surveyed \compcert's TCB \cite{Monniaux_Boulme_ESOP22}, both in the ``official'' releases and in our own ``chamois'' fork.\footnote{\url{https://gricad-gitlab.univ-grenoble-alpes.fr/certicompil/Chamois-CompCert}}
The trusted computing base can be extended in many ways, the most obvious of which is by adding axioms.

Axioms can be used to declare the existence of functions that are, through extraction directives, instantiated with OCaml code. In \compcert, this is in particular the case for many very short OCaml functions that look up the value of command-line flags. Just checking if an optimization or security feature should be active or not entails declaring an axiom that there exists such a Boolean flag, and then adding an extraction directive that explains how this flag may be obtained with a short bit of OCaml code. This usage does not increase the TCB.

In our work, the only axioms that we add, except for those dealing with command-line options, are the existence of a ``canary value'', which is only known when running the compiled program, as well as of processor-specific pointer encoding and decoding functions such that encoding then decoding a pointer yields that pointer. We will discuss this in more detail in \S\ref{sec:PAC_implementation}.

We implemented our work on top of the ``chamois'' version (\chamois{c145bbc5}{c145bbc5448d6ea30ccdff72ea57b23217f8215d}),%
\footnote{A frozen version is archived on Zenodo ({\small\texttt{\href{https://doi.org/10.5281/zenodo.10341149}{10341149}}}).}
proving our passes sound with no admitted results. The two new passes and the modifications to the stacking pass that we describe in this paper should however be usable with very minimal changes with the ``official'' version.
Options \texttt{-ftailrec}, \texttt{-fstack{\chyphen}protector}, \texttt{-fretaddr{\chyphen}pac}, respectively turn on tail recursion optimization (this also needs \texttt{-ftailcalls} so that tail recursive calls are exposed as tail calls), ``canary'' stack protectors and pointer authentication for return addresses. All these options are on by default and may be turned off by, e.g., \texttt{-fno{\chyphen}tailrec}. Finally, \texttt{-fstack{\chyphen}protector{\chyphen}all} in addition to \texttt{-fstack{\chyphen}protector} forces protection on all functions, not just ``vulnerable'' ones.

\section{Semantic Preliminaries: Undefined Behavior and Refinement}
We shall begin with considerations about how to model memory and undefined behavior.

The semantics of the intermediate and final representations in {\compcert} are defined as deterministic transition systems where the program moves from a state to another state.%
\footnote{We shall see later in this section how some constrained form of nondeterminism can still be expressed in such semantics. See \cite{DBLP:journals/pacmpl/ChappeHHZZ23} for more general views on modeling nondeterminism in Coq formalizations of semantics.}
The state, depending on which representation is used, may include certain elements such as the values of pseudo-registers or actual CPU registers, a more or less abstract vision of the call stacks, and a vision of memory.
A transition may or may not emit externally observable events. Externally observable events includes reads and writes to volatile variables and calls to external functions.
Various forms of simulations are used, and they always must preserve the \emph{trace of externally observable events}.

In contrast, the values inside registers, memory, etc. are deemed internal issues of the program and they may or may not be preserved by transformations and optimizations: most operations generate the empty trace of events~$\epsilon_0$.
This reflects the fact that the C standard mandates that the program behaves as though certain variables contain certain values, but does not mandate that these should be actually stored in memory at any specific location or even stored at all.

Programs transformations in {\compcert} are proved correct by exhibiting \emph{forward simulation} relations \cite{DBLP:journals/jar/Leroy09}.
Consider a transformation from a language $L$ to a language $L'$, with semantics respectively defined as transition systems over state spaces $\Sigma$ and $\Sigma'$. A ``match'' relation $M \subseteq \Sigma \times \Sigma'$ relates states of the program before and after the transformation.
$\Sigma$ and $\Sigma'$ are equipped with deterministic transition relations $\rightarrow$ and $\rightarrow'$: $a \rightarrow_t b$ means that state $a$ steps to  state $b$ emitting an event trace $t$ (in many cases, $t$ is the empty trace $\epsilon_0$).
The simplest case of simulation is \emph{lockstep}: if $\sigma_1,\sigma_2 \in \Sigma$ are such that $\sigma_1 \rightarrow_t \sigma_2$, and $\sigma'_1 \in \Sigma'$ is such that $M(\sigma_1,\sigma'_1)$ then there exists $\sigma'_2$ such that $M(\sigma_2,\sigma'_2)$ and $\sigma'_1 \rightarrow'_t \sigma'_2$.
More complex cases of simulation replace several steps by several steps.%
\footnote{See \module{common}{Smallstep}.}

One common way to model memory in the C programming language is as a set of disjoint memory blocks; a pointer consists in a block identifier and an offset within the block.%
\footnote{The C standard states that pointers point into \emph{objects}, and that testing the ordering of pointers into two different objects has undefined behavior \cite[\S 6.5.8-5]{C17}. It is however possible to test for (dis)equality pointers to different objects \cite[\S 6.5.9-6]{C17}; this suggests a modeling of pointers as the pair of an object identifier and a location within the object. In the Frama-C's \emph{typed memory model}, an address if a pair of base and offet, both integers~\cite[Sec.~3.7]{Frama-C_WP}. In Astrée, the memory model considers disjoint blocks of bytes, each representing a toplevel C object \cite{mine:hal-00136650}.}
This is the basis of the formalization of memory in {\compcert}, which considers blocks made of bytes~\cite{DBLP:journals/jar/LeroyB08,leroy:hal-00703441}.%
\footnote{Actually, the values stored in individual memory locations in {\compcert} cannot be exactly bytes. {\compcert}'s semantics allow for an unbounded number of memory blocks to be created without failure within a program execution, thus it is impossible to map pointers to true 4-byte or 8-byte words, which allow finitely many different values.
  For the purpose of simplicity, we will call ``bytes'' values that for the purpose of address computations take up one byte, but that may belong to an infinite type.}
This model was reused for the Vellvm formalization of the {\llvm} intermediate representation \cite[Sec.~6.3]{Zhao_PhD2013}.
% \cite{DBLP:conf/icfem/BlazyL05}.

It is possible, through pointer arithmetic, to move from any location in a block to any location in the same block, but not from a block to another; and ordered comparisons between pointers is defined only if they reside within the same block%
\footnote{In C, it is allowed to compute a pointer just past the end of an allocated block, so that loops such as \lstinline|for (int i=0; i<n; i++) *(t++)=42;| have defined behavior. Such a pointer may be compared to pointers within the block, but not be used for accessing data. For the sake of simplicity, we will leave out pointers just past the block from the discussion.}
(equality tests, however, are always defined). Each global or local variable resides in its own block, as well as each block allocated through \lstinline|malloc()|. A few examples:

\begin{lstlisting}
int a[3], b[3];
int i = b > a;  //undefined behavior
int j = b == a; //defined, yields false
int *a1 = a+1;  //defined, points to a[1]
\end{lstlisting}

In other words, in that model, a memory address consists of a block identifier $b$ and an offset $o$ such that $0 \leq o < l$ where $l$ is the length of the block. Pointer arithmetic modifies the offset while leaving the block identifier unchanged.

The C standard lists many cases of \emph{undefined behavior}. According to the standard, if a program executes one instruction with undefined behavior (such as integer division by 0 or out-of-bound memory load from an array), then anything can happen (actually, the whole execution trace has undefined behavior).

Depending which of {\compcert}'s formal semantics is concerned, undefined behavior is modeled either as an explicit ``stuck'' or ``none'' condition terminating the evaluation, or implicitly as the absence of a successor state in the transition relation. The provision in the C standard allowing execution to continue arbitrarily after reaching undefined behavior is reflected by the simulation relations, which states that if a sequence of steps is executed in the pre-transformation program, then a matching sequence of steps is to be taken in the transformed program, but does not put any constraints on the case where the pre-transformation program cannot take a step due to being stuck.

In {\compcert}, some forms of undefined behavior, instead of stopping program execution, are deemed to evaluate to a special {\Vundef} value:
all local variables contain {\Vundef} at function entry, all bytes are ``undefined'' just after memory allocation, etc.
Furthermore, arithmetic operations are strict with respect to {\Vundef}: $\Vundef + 42 = \Vundef$, etc.

{\Vundef} models some form of ``lazy'' undefined behavior, which simply propagates in the values instead of stopping program execution. Note, however, that {\Vundef} as an operand will stop program execution in the semantics when certain instructions are used. This is in particular the case of conditions: branching on a condition evaluating to {\Vundef}, instead of true or false, aborts execution. This is because one cannot define a unique next instruction in this case, which is required by the semantics.

It is always possible, during code transformation phases, to \emph{refine} a case where execution is stuck into a case where a variable gets assigned {\Vundef} (but not the converse). This is for instance the case of integer division: division by zero has undefined behavior at the C level and stops execution, but it is semantically possible at any point in the compilation chain to replace this division operator by another that, instead of not returning a value on division by zero, returns~{\Vundef}. This models the fact that, on many current processor architectures, division by zero does not trap. Such a transformation may, in turn, allow further optimizations.

Consider the following program, assuming \lstinline|x| is dead at the end of the code and its final value does not matter. One cannot move the division instruction before the test if it has ``stuck'' semantics on division by zero unless one can prove that division by zero never occurs:
\begin{lstlisting}
instructions modifying neither a nor b
if (condition) {
  x = a/b;
  y = x+3;
}
\end{lstlisting}
Yet, if we refine the division operator to another one that returns {\Vundef} on division by zero, we can, for optimization purposes, turn this program into:
\begin{lstlisting}
x = a/b;
instructions modifying neither a nor b
if (condition) {
  y = x+3;
}
\end{lstlisting}
This may be advantageous for efficiency: a division instruction may take time and it may be preferable to anticipate it. An out-of-order processor will do so on its own, but an in-order processor needs the compiler to anticipate the instruction.

The C standard also mentions \emph{unspecified} behavior, which does not terminate execution. Some operations may yield an unspecified value in some circumstances. The compiler is allowed to replace that ``unspecified'' value by an arbitrary actual value. Unspecified behaviors may also be modeled by {\Vundef}.
In other words, \emph{{\Vundef} can be used inside the deterministic semantics to model some form of nondeterminism}.

Program simulations are allowed to replace {\Vundef} by any value at any point. This is formalized by a \emph{definedness} partial ordering on values, where {\Vundef} is the bottom element and all other values are deemed greater than it (and not comparable with each other). This ordering extends pointwise to orderings on lists of values, memories, etc.
It is always possible, at any point, to \emph{refine} {\Vundef} into any other value.
Semantics for all program constructs are expected to be monotone with respect to this ordering: if some construct returns a defined value $v$ even if some of its operands are {\Vundef}, then, \emph{a fortiori}, it should return $v$ if some of these {\Vundef} are replaced by normal, defined, values.

\section{Memory Extensions and Canaries}
Stack canaries are countermeasures against buffer overflow attacks on variables, especially arrays, allocated on the stack. We shall first see how they operate, and then how they can be inserted inside a verified compilation chain.

\subsection{Canaries}\label{sec:canaries}
\begin{figure*}
\begin{center}\smallfigurefontsize
 \begin{tabular}{|p{0.2\textwidth}|p{0.3\textwidth}|l|l|l|}
   \hline
     optional register saves, spills\dots
   & stack-allocated local variables (including arrays, structures)
   & optional padding
   & \includegraphics[height=2em]{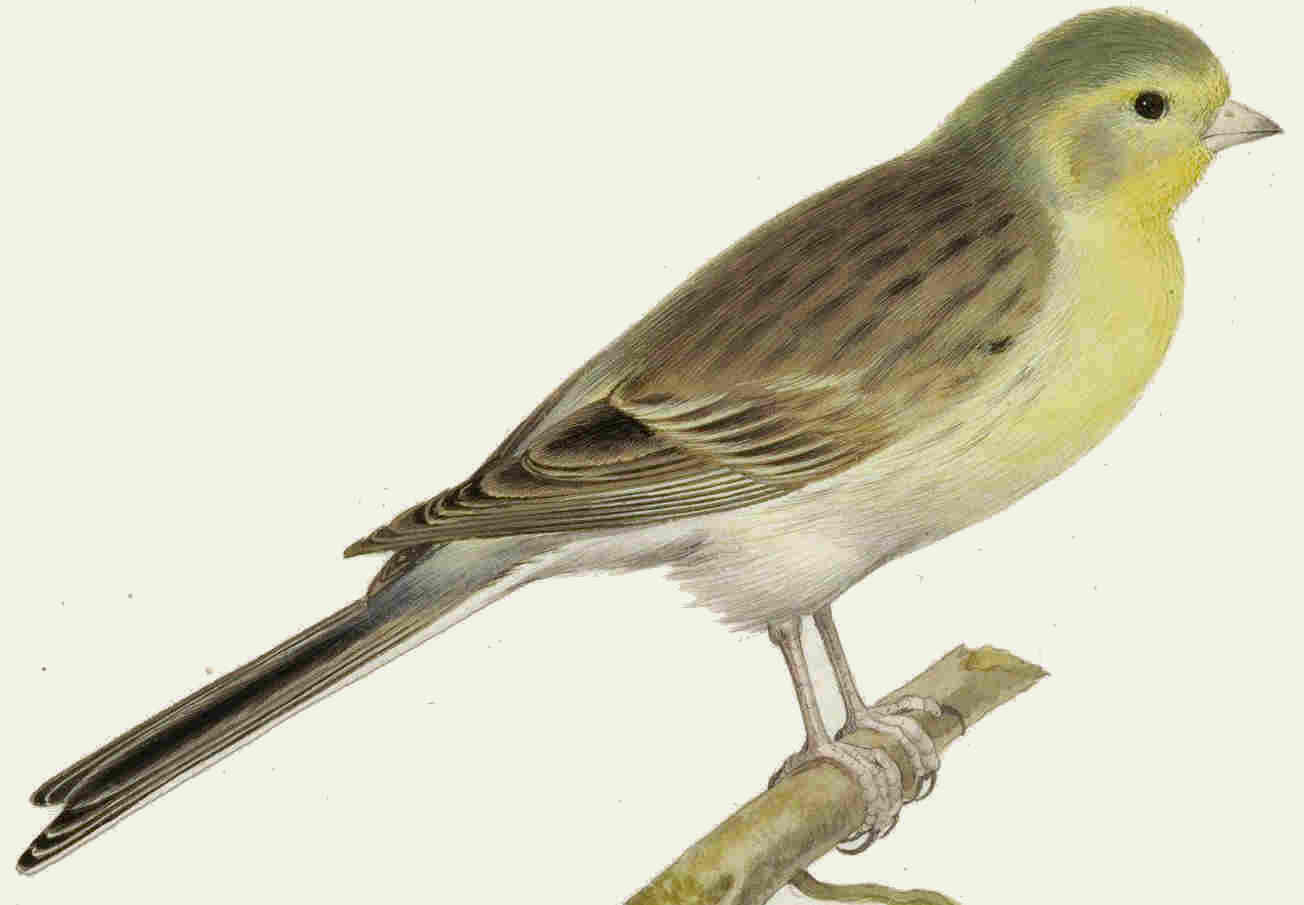}
   & return address\\
\hline
\end{tabular}
\end{center}
\Description[Stack frame completed with canary]{A modified stack frame: the original stack frame is followed by optional padding then the canary}
\caption{Stack frame with canary on x86 (see \texttt{Stacklayout.v} for the general layout; the canary is appended to the stack-allocated variable zone)}
\label{fig:stackframe}
\end{figure*}

It is well-known that a major cause of security vulnerabilities in programs in the C language (or any language based on C, such as C++, unless pointer use is restricted by programming guidelines) is the fact that memory accesses are never checked for validity. In particular, the language does not check if accesses to an array occur within its bounds (an array is considered as a pointer to its first element, and the length information is not carried).

The usual way of compiling C-like languages is to have functions allocate a \emph{stack frame} upon entry and deallocate it upon exit. The stack frame contains local variables (including, most interestingly, local arrays and structures handled through pointers), spilled register values (a register is \emph{spilled} to memory when not enough registers are available to store all live values), \emph{callee-saved} and \emph{caller-saved} register values (those that must be saved across function calls, depending on who is responsible for saving them), and, most importantly, the \emph{return address}, that is, the address in the code to which the program will branch when returning from the function. The stack frame is generally arranged in descending addresses (the stack frame of a function has lower addresses than that of its caller).

The position of the return address in the stack frame depends on the architecture and, possibly, on the compiler. In stack-oriented CISC architectures, such as 680x0 and x86, the return address is pushed on the stack during procedure calls, and then the stack frame is created below it (Fig.~\ref{fig:stackframe}). The return address thus resides at the end of the stack frame, after the local variables.
It is then possible to overwrite this return address by writing past the end of a local array, a not uncommon programming mistake.
In RISC processors, the return address is typically passed in a register (whether a special register, or a register defined by convention to contain the return address in the \emph{application binary interface}). In the case of leaf functions, that is, functions making no function calls, one can simply keep the return address in this register (\emph{leaf function optimization}). However, if function calls are made, the calls will overwrite this register, and thus the return address must be saved to the stack frame. Some compilers opt to save it at the beginning of the stack frame, which avoids making it vulnerable to writes past the end of the arrays. However, in this setting, the return address of the caller function, which resides in the stack frame of the caller function in memory (thus after the stack frame of the current function) may be overwritten by a write past the end of the array.

To summarize, if an array or structure is allocated on the stack and the function, for some reason, writes past the end of that array (or calls a function, such as a string manipulating function, that writes past the end of that array), it is possible to overwrite the return address of the function (or, depending on the architecture and compiler peculiarities, the return address of the caller function). In most cases, this will just result in the program crashing upon function return, because the address that will get loaded will contain garbage data that will not form a valid code pointer. If, however, the data written there is under the control of a malicious party, then it may point to executable code, which the program will branch to upon function exit. This is the infamous \emph{stack smashing} attack~\cite{smash_stack}.

Consider for instance the following program:%
\footnote{We store the loop index in an array to force {\gcc} to store it in memory before \lstinline|t[]|. If is is a scalar integer, {\gcc} stores it after the array, and the buffer overflow replaces it by the 32 low order bits of the address of \lstinline|quack()|, which is typically higher than \lstinline|i| and terminates the loop.}
\begin{lstlisting}
#include <stdio.h>
#include <stdlib.h>

void quack(void) {
  puts("quack"); exit(1);
}

void* vulnerable(int i, void* stuff) {
  void *t[10]; int j[1]; 
  for(j[0]=0; j[0]<i; j[0]++)
    t[j[0]] = stuff;
  return t[9];
}

void attack(int x) {
  vulnerable(x, (void*) &quack);
}

int main(int argc, char **argv) {
  attack(atoi(argv[1]));
  puts("normal");
  return 0;
}
\end{lstlisting}. If compiled with ``official'' CompCert \official{c7b27e3}{c7b27e30a9057c14d5a02a9f9aa3d8f230715eb0} on x86-64 and run as \verb+./canary 9+ (or any number $\leq 10$), the program prints \verb+normal+. However, if run as \verb+./canary 12+, it prints \verb+quack+.
What happens is that \lstinline|t[10]| is the location for the return address of the \lstinline|vulnerable| function, which gets overwritten with the address of the \lstinline|quack()| function.
With the protection we implemented into {\compcert}, which we shall describe below, it instead prints the following message before aborting:
\begin{verbatim}
*** stack smashing detected ***: terminated
\end{verbatim}
The same applies to {\gcc}-9.4.0 whether one uses \texttt{-fstack{\chyphen}protector} or \texttt{-fno{\chyphen}stack{\chyphen}protector}.

The above simple example is unrealistic, because it assumes the attacker is within the same program. More realistic attacks involve exploiting bugs in the targeted application or system software.
In the simplest cases, the malicious party provides some incorrectly formatted data, which triggers some unforeseen and undefined behavior of the function that causes it to overflow some data structure and overwrite the return address to a value pointing into malicious data stored on the stack (or in blocks allocated by \lstinline|malloc()|), which is then executed as code. Of course, this assumes that the attacker is able to predict the addresses at which this data is loaded, but this is often the case assuming the attacker has access to the binary that is run on the target platform (which is the case if it is a standard program), unless ASLR is used.%
\footnote{%
\emph{Address space layout randomization} (ASLR), implemented in e.g. Linux and MacOS, changes the base addresses of various structures at every execution, so as to make them not predictable by the attacker. As of 2023, current Linux distributions, by default, turn ASLR on and make the stack and other memory data non executable. ASLR can be turned off for debugging purposes.}

Such simple attacks, which rely on \emph{code injection} (having data from the outside pass off as code) have long been thwarted by making the stack (as well as memory allocated by  \lstinline|malloc()|)) nonexecutable by default, but this requires a processor capable of differentiating executable memory from nonexecutable memory (all current application-grade processors with a memory management unit are capable of this, but not necessarily processors for simple devices).%
\footnote{Memory can be made executable through system calls, for instance to implement just-in-time compilation.}
\emph{Return into libc} eschews code injection by returning to a suitable function inside the standard C library~\cite{Solar_Designer_1997}.
\emph{Return-oriented programming} (ROP)~\cite{ROP} greatly extends this idea of using existing code inside the targeted application. It works by writing a succession of return addresses to code fragments, also known as \emph{gadgets}, already present inside the program code. The thesis behind ROP is that in any sufficiently large code base, one can always find a succession of gadgets that allow executing sufficiently arbitrary code to gain control of the application. 
However, all such kinds of attack may often be stopped by using \emph{canaries}.

A \emph{canary} or, more formally, a \emph{stack protector}, is a special value written at function entry inside the stack frame, past any structure that may be adressed through pointers but before the return address, and checked to be still present at function exit~\cite{Cowan_et_al_821514}. If this value is not longer present, then stack smashing is detected and the program terminates.
The idea is that in most cases, buffer overruns occur in contiguous areas and such an overrun, if it can reach the return address, must overwrite the canary.%
\footnote{This system is named after the use of canaries or other small birds inside coal mines as a warning system. The birds would become sick before the miners when breathing noxious gases, especially carbon monoxide, and thus would warn the miners. In our case, the canary dies from incorrect memory accesses as a warning to the program before these can be exploited by the malicious party.}
The value of the canary is typically chosen randomly at program start so that it may not be guessed by the malicious party.

A compiler may implement canaries by%
\begin{itemize}
  \item
    adding to function prologue the sequence: read the canary, write it to the stackframe (optionally, clear the register containing the canary)
  \item
    adding to function epilogue the sequence: read the canary, read the value stored in the canary location in the stack frame, branch to an error crash sequence if the two values do not match
\end{itemize}
The informal argument for this code transformation working is that if the program executed correctly prior to the transformation (recall that writing to the canary counts as undefined behavior thus incorrect execution), then this execution is preserved after the transformation.

One can choose to install canaries on all functions or, to avoid unnecessary slowdown, only on ``vulnerable'' functions, according to some heuristic for vulnerability: for instance, functions that have arrays inside the stack frame.

\subsection{Formal Semantics and Verified Insertion}
Let us now see how the above informal description of stack canaries can be made to work inside a verified compiler.
Assume now that stack-allocated data is semantically modeled as a single block, as is the case in {\compcert}.%
\footnote{\compcert, early in the compilation phases, allocates into the stack frame, considered as a single memory block, all local variables whose address is taken, including structures and arrays; the other local variables are allocated to pseudo-registers. Arguably, one would have made the choice of keeping each variable separate, which would allow, for instance, inserting canaries between variables on the stack and not just at the end of that single block.}

The informal argument above translates into the statement that if the program executes correctly with stack frames of certain lengths, then it executes correctly with longer stack frames.
Note that the program with canaries and thus longer stack frames has \emph{more} legal executions than the original program. This is an effect of modeling stack frames, including the canary, as a single memory block.
Imagine for instance that the original function has $32$ bytes of stack-allocated arrays, and some instruction accesses these at offset $32$; this program evidently has undefined behavior because that access is outside of those arrays, and execution will terminate in the semantics of the intermediate representation. Suppose now we add a canary to that function, inside the same block, at offset 32. Then, reading from that location will just return the canary value.
An execution with a write to offset 32 gets stuck at the time of the write in the original program, but will proceed in the transformed program, until the canary is checked; thus the execution will last longer.

In addition to executions going further, extending stack frames may also turn values from {\Vundef} to defined. Consider an order pointer comparison $p < p+34$ between $p$, the start of the stack frame of length $32$, and $p+34$. Because $p+34$ points outside of the block, this comparison returns {\Vundef}. However, if we extend the stack frame with a canary of size $8$, this comparison becomes defined and returns true, as expected.

{\compcert}'s notion of \emph{memory extension} \cite[\S5.2]{DBLP:journals/jar/LeroyB08} captures the points we have expressed before. A memory $m'$ extends a memory $m$ if $m'$ contains all the blocks of positive size in $m$, with lengths greater than or equal to their length in $m$ (it may also contain other blocks).
If a memory location exists in $m$, then it must exist in $m'$ and its contents must be more defined than in $m$, that is, if the contents at a location in $m$ are defined, then the same contents must appear at that location in $m'$.
Calls to external functions are axiomatized in a way that guarantees that if a function call with arguments $a$ and memory $m_1$ succeeds and returns $v$, with modified memory $m_2$, then if $m'_1$ extends $m_1$ and arguments $a'$ are more defined than $a$, the same function returns $v'$ more defined than $v$ in a memory state $m'_2$ that extends $m_2$.

The proof of correctness of the canary installation phase relies on matching execution states through a memory extension and ``more defined'' relations. Furthermore, for every function on the call stack, if that function has a canary, then that canary must be in the right position.
In addition, we state that in the original program, the stack frame block has the same size as the stackframe (no extra space) and that in the transformed program the extra stacksize space for padding and canary is actually allocated.

\subsection{Implementation and Proof}
The Coq implementation of the transformation in {\compcert} is 97 lines long and the proof is 1689 lines long. It operates on the \verb+RTL+ representation, which is a form of ``3-address code'' where instructions are arithmetic operations, loads, stores, conditional branches, builtin calls, procedure calls, tail calls, and branches through jump tables.

We formalized the ``get canary'' operation as an operator returning an uninterpreted integer constant ($c_{32}$ or $c_{64}$ depending on whether the canary is $32$- or $64$-bit on the target platform), which gets expanded in a platform-dependent way when producing assembly (the method for retrieving the canary depends on the architecture and the operating system; for instance on x86 Linux it resides at a special offset within thread-local storage, which is to be accessed using specific segment registers).

The transformation pass identifies functions that must be fitted with canaries, computes the offset at which the canary is inserted in the stack frame (some padding may be needed to satisfy alignment constraints) and the size of the extended stack frame, inserts a ``load canary, save it to the appropriate location'' sequence at function entry, adds a call to a special function \lstinline|stack_chk_fail|, and then, for each return or tail call (see~\ref{sec:tailrec}), precedes these call with a ``load canary, load saved value, compare them, and branch to the special function call if they differ'' sequence.

We do not add axioms to those already present in {\compcert}, except those necessary to interface with new command-line options, and one that states that there exists a canary value: thus the impact to the TCB is null.

For each platform that we support (currently, only Linux on x86, x86-64, RISC-V, AArch64 and ARM), we provide specific code for retrieving the canary.%
\footnote{We have not found authoritative documentation on how to retrieve the canary and thus follow what {\gcc} and {\clang} do, though not necessarily with the same instruction sequences. In particular, {\gcc} will erase the register in which the canary is used, but {\compcert}'s register allocation will optimize out erasure instructions. Fixing this is left for future work.}

% (stack: list stackframe) (**r call stack *)
%             (f: function)            (**r current function *)
%             (sp: val)                (**r stack pointer *)
%             (pc: node)               (**r current program point in [c] *)
%             (rs: regset)             (**r register state *)
%             (m: mem)
The proof of correctness is, as usual in {\compcert}, proof that the transformed program simulates the original program.
This proof relies on a matching invariant $\sim$ between the states of the original and the transformed program.

At \verb+RTL+ level, a regular state $\sigma$ is a tuple $(\textit{stack},f,\textit{sp},\textit{pc},r,m)$ where \textit{stack} is the abstract call stack, $f$ is the function currently executed, \textit{sp} is the address of the current concrete stack frame in memory, \textit{pc} is the number of the current instruction under execution within $f$, \textit{rs} is the state of the bank of pseudo-registers, and \textit{m} is the memory; there are also \emph{call states} for when a function is just about to start to be executed, and \emph{return states} for when a function is just about to return.
%      forall (res: reg)            (**r where to store the result *)
%             (f: function)         (**r calling function *)
%             (sp: val)             (**r stack pointer in calling function *)
%             (pc: node)            (**r program point in calling function *)
%             (rs: regset),
The abstract call stack is a possibly empty list of abstract stack frames, with the immediate caller of the function currently under execution at the start of the list. Each abstract stack frame is a tuple $(\textit{ret},f,\textit{stack},f,\textit{sp},\textit{pc},r)$ where \textit{ret} is the number of the pseudo register in the caller function where to store the return value, and $f$, \textit{sp}, \textit{pc} and $r$ have the same meaning as for the function currently executed, except that they pertain to the function whose execution was stopped when it called another; thus \textit{pc} points to the instruction to which to return.

We use a ``plus'' simulation, meaning that if there are states $\sigma_o$ and $\sigma_o'$ in the original program such that it may take a step $\sigma_o \rightarrow \sigma_o'$, and $\sigma_o \sim \sigma_t$, then there must exist $\sigma_t'$ and a nonempty sequence of steps  $\sigma_t \rightarrow \dots \rightarrow \sigma_t'$.
The reason we need a nonempty sequence of steps in the transformed program, and not just a single one, is that the prologue, epilogue, and tail call instructions, which take one single step in the original program, need several steps to retrieve and store or check the canary. The other operations are handled in lock-step.

The matching invariant $\sigma_o \sim \sigma_t$ essentially states that in the current function, as well as all stack frames in the abstract stack frame:
\begin{itemize}
\item the memory $m_t$ of the transformed program \emph{extends} that of the original program, $m_o$
\item the canary value is stored at the appropriate position in the concrete stack frame block in $m_t$, with appropriate permissions
\item the canary value is inaccessible in $m_o$, meaning that it cannot be read or written in this memory.
\end{itemize}

The formal proof essentially argues that if a store operation or a call to an external function succeeded in the original memory $m_o$, where the canary value is inaccessible, then it cannot attempt touching the canary value (otherwise execution would have stopped, because the canary location is inaccessible in that memory), thus these operations never touch the canary value in~$m_t$.
In addition to that, it argues that the prologue sequence actually loads the canary at the correct location in the concrete stack frame block (necessary for the matching invariant), and that the function return and tail call sequences never branch to the ``stack smashing detected'' procedure.
A tedious part of the proof shows that the prologue, tail call and return sequences are correctly added to the function without clobbering other operations.

Note that we do not prove that canaries constitute an \emph{adequate} security measure, that is, that the program will go to the ``stack smashing detected'' procedure if stack smashing happens in the original program. It is in fact possible to prove that in the transformed program, the call to ``stack smashing detected'' is unreachable using normal execution semantics.
The definition of adequation is delicate. Ideally, we would like to ensure that the function would return to its caller and not elsewhere; but a canary does not ensure that, because it is possible for the attack to ``jump'' over it.
The design of a weaker adequation property and its proof are left to future work.

\section{Memory Injections and Tail Recursion Elimination}
In functional programming languages, it is commonplace to program iterations through \emph{tail recursion}, that is, calling the function itself as its final action before exiting. It is in fact expected, in such languages, that the compiler will turn tail recursive calls into loops.
We shall see here how such an optimization can be formally proved.

\subsection{Tail Recursion Optimization}
\label{sec:tailrec}
Consider the following function:
\begin{lstlisting}
int f(int x) {
  instructions
  return g(x+5);
}
\end{lstlisting}
The normal way of executing \lstinline|f| would be to save the return address of \lstinline|f| (the address to which \lstinline|f| must return) and create a stack frame for \lstinline|f|, save callee-saved registers, compute $x+5$, then call \lstinline|g|, restore callee-saved registers, destroy the stack frame for \lstinline|f|, restore the return address of \lstinline|f|, and finally return from \lstinline|f| (there are some architectural variations whether the return address is saved before or after the stack frame is created, and restored before or after the stack frame is destroyed).

Remark that the value returned by \lstinline|g| is directly used as the return value for \lstinline|f| with the same memory state. One could instead destruct the stack frame for \lstinline|f|, restore the return address, and jump to \lstinline|g|. Then \lstinline|g| gets executed and returns to whoever called \lstinline|f|. This avoids accumulating useless frames on the stack. This optimization is known as \emph{tail call optimization}.

This optimization is correct only if no pointers into the local variables of \lstinline|f| escape to \lstinline|g|. A sufficient condition for this is that \lstinline|f| has no stack-allocated local variables.%
\footnote{Because this constraint is appreciated prior to register allocation, the stack-allocated local variables it takes into account are arrays, structures, and scalar variables whose address is taken. Variables that are spilled to stack locations because of register allocation do not count.}
This is how {\compcert}'s existing tail call elimination pass \texttt{Tailcall} operates; we do not relax this restriction.%
\footnote{The \texttt{Tailcall} pass operates over the \texttt{RTL} representation. Tail calls are however available in previous intermediate representations starting from \texttt{Cminor}, but the C frontend does not generate them. They may however be generated by alternate frontends~\cite[\S 7.3.4]{DBLP:phd/hal/Dargaye09}.}

A common case for tail calls is \emph{tail recursion}: the tail call is made to the same function. If the above tail call elimination is used, a tail call from a function $f$ to itself amounts to restoring callee-save registers, destroying the stack frame, jumping to the head of the function, creating a stack frame of the same exact size and saving the callee-save registers (in the same positions). This seems like a waste of time; we would expect \emph{tail recursion elimination}, which turns tail recursion into a loop that keeps the stack frame as is.

For instance, this function:
\begin{lstlisting}
int descend(tree *t, char *u, int n) {
  if (n <= 0) return t;
  if (*u)
    return descend(t->right, u+1, n-1);
  else
    return descend(t->left,  u+1, n-1);
}  
\end{lstlisting}
can be turned into:%
\footnote{This could be more elegantly be written using a \lstinline|while| loop, but the way optimizations work over a control-flow graph is to insert \lstinline|goto|'s.}
\begin{lstlisting}
int descend(tree *t, char *u, int n) {
  loop: if (n <= 0) return t;
        if (*u) { t=t->right; u++; n--;
                  goto loop; }
        else    { t=t->left;  u++; n--;
                  goto loop; }
}  
\end{lstlisting}

Tail recursion elimination improves on tail call elimination by witnessing that destroying an ``old'' stack frame just to create a ``new'' one of the same length is useless and one should instead reuse the same stack frame.

\subsection{Formal Semantics and Verified Optimization}
Again, we must find a formal counterpart to the informal argument that one can freely reuse the existing stack frame.
The proof of this optimization relies on the notion of memory injection, a generalization of extensions.

A \emph{memory injection} \cite[\S5.4]{DBLP:journals/jar/LeroyB08}
% \cite{DBLP:conf/itp/BessonBW15}
consists in mapping all currently allocated memory blocks in the original program to sub-segments of memory blocks in the transformed program. Injections are notably used for proving the correctness of the \emph{inlining} transformation, where a call from a function \lstinline|f| to a function \lstinline|g| is replaced by the code for \lstinline|g| (with suitable parameter passing and register renaming): the stack frame of the combined function contains the stack frames for \lstinline|f| and \lstinline|g|.

A memory injection $\mu$ maps a block $b$ either to nothing (see below for one use for this), or to a block $b'$ and offset $o'$. A byte at offset $o$ in block $b$ is mapped to offset $o+o'$ in block $b'$.
Memory injections thus constrain the semantics of program operations: they must commute with injections.
Adding an integer $i$ to a pointer $(b,o)$ commutes with injections: $(b,o)+i = (b,o+i)$; $\mu(b,o+i) = (b',o'+o+i) = \mu(b,o)+i$.
In contrast, checking that specific bits in the offset have certain values is an operation that does not commute with injections.
The formal requirement that program semantics commute with memory injections corresponds, partially, to the informal idea that ``program behavior must not depend on the addresses to which variables are allocated''.%
\footnote{In C, comparing for order pointers to distinct variables, living in distinct memory blocks, has undefined behavior.}
The same applies to the semantics of external calls.

For the proof of our tail recursion elimination phase, the injection maps the identifier of the ``new'' stack frame in the original program to the identifier in the new program to which the ``old'' stack frame was mapped, while the ``old'' stack frame maps to nothing.

In practice, for {\compcert}, for each function we identify tail calls to the same function, and for each of these calls we replace them by
\begin{itemize}
\item a sequence of copies of the effective arguments to the tail call to temporary variables
\item a sequence of copies from these temporary variables to the positional arguments
\item an unconditional branch to the start of the function.
\end{itemize}
Note that this creates a loop that is amenable to any loop optimization in further passes.

Our copy sequences perform, in effect, a parallel assignment of effective to positional arguments. There are algorithms for generating minimal sequences of moves for such assignments. We however opted for simplicity. Further optimisations of assignments and register allocation anyway suppress most such moves in practice.

Again, what is done is a refinement, not an exact simulation. On normal function start, all pseudo-registers contain {\Vundef}. However, when a tail-recursion call is replaced by the above sequence, the pseudo-registers that are not positional arguments to the function retain their last value. This is not a problem, because the pseudo-registers at function entry in that case are ``more defined'' that the pseudo-registers at normal function entry, which satisfies the ``match'' relation.

\subsection{Implementation and Proof}
The Coq implementation of the transformation is 69 lines long (including some sanity check about instruction numbers that could be removed). Again, it operates on \verb+RTL+.
The proof of correctness is 2641 lines long, half of which are rather tedious proofs about inserting instructions at correct locations and preserving them during further insertions, and the other half the inductive invariants discussed above. The invariants and the proofs of memory injections were nontrivial.

Consider for instance this tail recursive implementation of factorial:
\begin{lstlisting}
static int fac_rec(int x, int acc) {
  if (x >= 1) return fac_rec(x-1, x*acc);
  else return acc;
}
int fac(int x) {
  return fac_rec(x, 1);
}
\end{lstlisting}

Our modified version of {\compcert} will turn the recursive call in \lstinline|fac_rec()| into a loop. Since \lstinline|fac_rec()| then becomes a non-recursive function callable only from one location, in \lstinline|fac()|, it gets inlined into \lstinline|fac()|. Since there are no more possible calls to \lstinline|fac_rec()| (recall that a \lstinline|static| function cannot be called from outside of the module), it is deemed useless and removed from the module.
The \lstinline|fac()| function is thus compiled to assembly code equivalent to (showing a \lstinline|while| loop instead of \lstinline|goto| for readability):
\begin{lstlisting}
int fac(int x) {
  int acc = 1;
  while(x >= 1) {
    acc = acc * x;
    x = x-1;
  }
  return acc;
}
\end{lstlisting}

Similarly, if compiling
\begin{lstlisting}
void twisted_quicksort(data x,
    data *A, int len) {
  if (len < 2) return;
  data pivot = A[len / 2];
  int i, j;
  for (i = 0, j = len - 1; ; i++, j--) {
    while (x*A[i] < x*pivot) i++;
    while (x*A[j] > x*pivot) j--;
    if (i >= j) break;
    data temp = A[i];
    A[i]     = A[j];
    A[j]     = temp;
  }
  twisted_quicksort(x, A, i);
  twisted_quicksort(x, A + i, len - i);
}
\end{lstlisting}
the second recursive call to \lstinline|quicksort()|, the tail one, gets compiled into a subtraction, a move and a jump to the head of the function.%
\footnote{%
This function, where \lstinline|data| can be instantiated to be \lstinline|uint32_t| or \lstinline|uint64_t|, sorts elements according to $a \prec_x b \iff (x.a) \bmod 2^n < (x.b) \bmod 2^n $; for odd $x$, $a \mapsto x.a$ is a permutation and thus this is a total ordering. We use it for evaluation purposes because it is possible to obtain different orderings on the same data by just changing~$x$.}

We do not add axioms to those already present in {\compcert}, except those necessary to interface with new command-line options, and thus the impact to the TCB is null.

\section{Pointer Authentication for Return Addresses}
\label{part:PAC} 
We shall now investigate how to implement in a formally verified manner another mechanism for thwarting stack smashing: pointer authentication (PAC), as offered by the ARM v8.3 architecture \cite{Qualcomm_ARM_PAC,PAC_iPhone_XS,PAC_xnu,Apple_PAC_LLVM}.

\subsection{Semantics}
The operating system, if it supports this functionality, sets up secret keys in the processor. The processor has two imp\-lemen\-ta\-tion-spe\-ci\-fic functions $E(p,m,k)$ (encoding) and $D(c,\allowbreak m,k)$ (decoding), where $p$ is a pointer, $m$ is a modifier, $k$ is a secret key, such that $D(E(p,m,k),m,k)=p$ for any $p,m,k$.
The modifier is an arbitrary value, the only requirement on it is that it should be the same for encoding and decoding; suitable examples include zero and the stack pointer (if one is sure that the value is decoded in the same function it was encoded in).
The idea is that it would be in practice impossible for a malicious party to guess $e$ so that $D(e,m,k)$ is a valid pointer.

Before storing a pointer $p$ to a location where it could be overwritten by vulnerable code, one applies $e=E(p,m,k)$, and apply $p'=D(e,m,k)$ to recover $p'=p$ before using it.
Then, even if the malicious party can overwrite this location with arbitrary data, it cannot freely redirect the pointer stored there.
If there were only $p$ and $k$, the malicious party could perhaps, by triggering some memory copy operation, have an otherwise valid pointer copied and reused in another context in which it was not intended to be used; the modifier is intended to make this harder.

In particular, there exist two instructions (\verb+paciasp+ and \verb+autiasp+) that respectively apply $r:=E(r,s,A)$ and $r:=D(r,s,A)$ where $r$ is the return address register and $s$ is the stack pointer.
The idea is that one applies \verb+paciasp+ before saving the return address to the stack, and \verb+autiasp+ before restoring it.
On earlier processors not supporting these two instructions, they are executed as ``no operation'', which amounts to saying that $E$ and $D$ are identity functions ($E(p,m,k)=p$ and $D(e,m,k)=e$).

We modified {\compcert}'s matching invariants that state that the return address is stored at a certain offset in the stack frame so that instead of storing it always in the clear it may be stored as its image by~$E$. As $E$ and $D$ are implementation-specific, we just axiomatize them as uninterpreted functions mapping values to value such that $D(E(p,m,k),m,k)=p$ if $p$ is a pointer.\footnote{%
  Such a symbolic view of encoding and decoding, when applied to encryption / decryption primitives, is known as the \emph{Dolev-Yao model} \cite{DBLP:journals/tit/DolevY83}.}

Now, $D$ and $E$ certainly do not commute with memory injections. They are indeed intended so that knowing that $e=E(p,m,k)$ does not help the malicious party forge $e'$ such that $e'=E(p',m,k)$ and $p'$ is a pointer of interest to the malicious party. However, since the handling of return addresses appears late in {\compcert}'s code transformation phases, this lack of commutation does not hinder proofs, since no further transformation phase relies on injections for its proof of correctness.

If we run the attack described in \S\ref{sec:canaries} without canaries but with pointer authentication, we obtain:
\begin{verbatim}
normal
Segmentation fault (core dumped)
\end{verbatim}
The attack modified the return address of \lstinline|main()|, but did not overwrite it with an authenticated pointer, thus the \verb+autiasp+ instruction yielded an incorrect pointer, which led to a segmentation fault when executing the return instruction.

Finally, the ARM 8.3 instruction set introduced the \verb+retaa+ instruction, which combines \verb+autiasp+ and the function return instruction. This instruction does not work on previous generations of processors; we thus generate it only if a certain option is passed.
For this, we extended the existing postpass optimizer introduced by \citet{Gourdin_AFADL21}:%
\footnote{This optimizer replaces combinations of machine instructions by a single instruction with identical semantics. It operates post all other optimizationpasses.
So far, it was used to replace load and store instructions of single values in consecutive memory locations by load and store instructions capable of moving multiple values. An untrusted oracle proposes a new instruction sequence; its result is validated by performing symbolic execution on both the original and modified sequences and checking the final symbolic results are identical, which implies they have identical semantics.} 
if the option is active, this pair of instructions is replaced by a single \verb+retaa+.

We had to deal with some slight semantic issue in order to do so, since this instruction is not exactly equivalent to the pair of instruction it replaces:
\verb+autiasp+ ``decodes'' the return address into the return address register, which is then used by the return instruction; \verb+retaa+ does not change the return address register.
The peephole optimizer verifies instruction replacement through symbolic execution and the two sequence of instructions must have identical final result, which is not the case here.
We worked around this difficulty by stating that both \verb+retaa+ and the normal return instruction leave an undefined value (\Vundef) in the return address register---this is acceptable since their actual behavior, which is to leave the value of that register unchanged, is a refinement of this.
Then, the two-sequence instruction \verb+autiasp+ plus return, and the single \verb+retaa+ instruction have the same semantics and the transformation is validated.

\subsection{Implementation and Proof}
\label{sec:PAC_implementation}
The implementation consists in a 54-line \texttt{PAC} module and 11 changes to invariants and proofs in the \texttt{Mach} and \texttt{Stacking\-proof} modules of \compcert, those that implement the storing or restoring of return addresses. We also added the \verb+autiasp+, \verb+paciasp+ and \verb+retaa+ instructions at the assembly level.

To each function is associated a pointer authentication kind.
For each architecture, we provide two functions: one (\verb+choose_pac_kind+) chooses, given a program function, the kind of pointer authentication used in this function; the other, (\verb+pointer_auth_encode+), encodes the pointer according to the authentication kind, the pointer, and the modifier. One must also provide a property that if both the pointer and the modifier are of the pointer type (that is, a 32-bit or 64-bit integer depending on the architecture), then the encrypted pointer belongs to that same type.

On all architectures except AArch64, the type of authentication kinds is a singleton (the only authentication kind is ``none''), the encoding function is the identity, and the lemma about the pointer type is trivial.

On AArch64, we implement two authentication modes: ``none'' and ``key A'' and thus the kind type has two constructors.%
\footnote{AArch64 also has a B key. We currently support only key A, since this is what {\gcc} uses.}
We add 6 axioms to {\compcert}.
One is extracted into an OCaml function that deals with the command-line option and selects the kind type accordingly. The kind also formally depends on the function under compilation, but we currently do not take it into account; we could for instance implement a command-line option disabling pointer authentication on leaf functions that do not restore the return address from memory. This function does not add to the TCB.

The 5 remaining axioms are the only genuine axioms added in this work. One states that there exists an encoding function for key A; another that there exists a decryption function for key A; another that encoding then decoding is the identity function; another that if the result of encoding with key A is {\Vundef}, then either the pointer or the modifier (or both) was {\Vundef}, and finally that the result of encoding a value with pointer type with key A also has pointer type.

Could these axioms introduce logical inconsistency? We show that this cannot be the case by showing that these axioms can be satisfied by instantiating the encoding and decoding functions for key A by the identity and that, in this case, the remaining axioms can actually be proved.

Could we do without adding axioms? Since ARM does not specify exactly the actual computations that the pointer encoding and decoding functions perform, and in fact reserves the right to implement them differently in different processors, it is inevitable that, whatever the formalization we choose, there is a point when we need to state as axioms that there exist some functions that the processor performs and that satisfy certain properties.

As for canaries, the correctness proof relies on a simulation between states in the original and transformed programs. In this case, however, we need a ``star'' simulation, where a succession of steps in the original program is simulated by a succession of steps in the transformed program.
In almost all cases, our proof is in fact lock-step (one step in the original matches one step in the transformed program), but, when tail recursion elimination is applied, we replace a sequence of two steps (entering the call state of the tail-called procedure and stepping from that call-state to the first normal state in the called procedure) by a sequence that copies parameter values to temporary pseudo-registers, then copies them into the parameter variables, and finally branches to the top of the function.

The matching invariant relies on a memory injection. In most cases, the construction of the injection follows both executions: if the original program creates a memory block $b_o$ and the transformed program creates a memory block $b_t$, then the injection maps $b_o$ to $b_t$ at offset $0$.
When it comes to the tail recursive call, the injection, at the end of the call, maps the new stack frame block $b'_o$ of the current function (the one that has been re-created after the old stack frame block $b_o$ was destroyed) to the current stack frame block $b_t$, while $b_o$, the old, destroyed stack frame block in the original programs, is removed from the map.

In order to show all the required properties for the injection to hold, we need an auxiliary invariant that, for the stack frame of the current function as well as all functions on the stack, that stack frame is ``weakly allocated'', that is, if one has permission to access a memory byte in the concrete stack frame block, then the offset $o$ of the byte in the block is such that $0 \leq o < s$ where $s$ is the size of the stack frame.%
\footnote{One would expect that the converse (a byte at address $(b,o)$ such that  $0 \leq o < s$ where $s$ is the size of a stack frame and $b$ is the number of the block for that stack frame is always accessible) would also be true. However such an invariant cannot be proved within {\compcert}, because the semantics of external calls is too lax. Specifically, the axiomatization of external calls allow the external function to free arbitrary memory locations, including the stack frames of caller functions.

Of course this does not make sense with respect to final assembly-level semantics: a function may only free its own stack frame. Even when considering exception handling or \lstinline|longjmp()|, one cannot transfer control to a function whose stackframe has been destroyed.}

There is also a significant amount of tedious proofs showing that the tail recursive call sequences are correctly placed, that their placement did not clobber other instructions, and that the parameter copies that they perform to and from temporaries truly place the parameter values into the correct pseudo-registers.

\section{Experimental Evaluation}
{\compcert}'s correctness proof states that if it succeeds in compiling a program, the formal semantics of the assembly code that it produces simulates the formal semantics of the C source code. It does not state that {\compcert} will necessarily succeed in compiling within reasonable time.

Furthermore, at the end of {\compcert}'s backend there are a 2-3 phases (depending on the target architecture) that are implemented in OCaml and are trusted to be correct, including the one that prints assembly code in text form. Bugs have been found in those passes through testing~\cite{Monniaux_Boulme_ESOP22}, as well as in the formal semantics given for (pseudo-)assembly instructions. Since, for pointer authentication, we added new instructions along with their semantics, and slightly modify the semantics of one existing instruction, testing is again needed.

To detect such bugs, we use the same test suite we used previously \cite{DBLP:conf/tap/MonniauxGBL23} in our continuous integration system and with which we identified code generation bugs in both our fork and the ``official'' releases of {\compcert}.
This test suite consists of the small test suite shipping with {\compcert} ``official'' releases, to which we added hundreds of randomly generated programs produced by 3 different generators (CSmith, YarpGen, CCG), and most of {\gcc}'s ``torture test'' suite.
In some cases, the resulting object code is run and must match a known result, in other cases one only checks that the compiler ran successfully.
We ran this test suite without encountering errors.

We also provide a larger test suite for benchmarking the speed of executable code, suitable for measuring the impact of optimizations and security features.%
\footnote{\url{https://gricad-gitlab.univ-grenoble-alpes.fr/certicompil/chamois-benchs}}
Because pointer authentication is only available on AArch64, we ran the test suite on a Raspberry Pi~3 equipped with an ARM Cortex A53 processor, which interprets pointer authentication as ``no operation''. Return address authentication and pointer authentication result in a $< 1\%$ slowdown.
The same applies to the Apple M1 processor, which does implement pointer authentication: the slowdown is too about 1\%. (The stack protector is not yet available on MacOS, and thus we did not time its impact.)

Stack canaries inserted only on ``vulnerable'' functions (functions with stack-allocated variables, register spills excluded) results in a $< 1\%$ slowdown, and a 5\% slowdown if applied to all functions. This vindicates the strategy, already implemented in {\gcc} and {\llvm}, to only apply them to vulnerable functions. Future work should include a better strategy for designating which functions are at risk (pointer arithmetic / array accesses, pointers to local variables passed to other functions?).

Our purpose in this work was to implement standard optimizations and security features and to show how they can be proved to be semantically correct, not to show that these standard optimizations bring large speed improvements. Improvements to tail recursion affect only a narrow category of functions: rather small functions with tail calls. Let us see two examples of these.

The use case for tail recursion elimination for functions traversing data structures can be illustrated by:
\begin{lstlisting}
int last(struct list *p) {
  if (p->next == NULL) return p->val;
  else return last(p->next);
}
\end{lstlisting}

% gcc 18.911s
% gcc -O 11.222s
% gcc -O2 5.579s
% ccomp -fno-tailcalls 11.529s
% -ftailcalls -fno-tailrec 6.647s
% -ftailcalls -ftailrec 0m5.695

On a RISC-V ``Rocket'' core implemented in a FPGA, this procedure ran 100 times on a 100000-element list takes 6.6~s using {\compcert}'s tail call elimination, and only 5.7~s (the same timing as \verb+gcc -O2+) if using tail recursion elimination, a 14\% reduction. (Disabling tail call optimization results in 11.5~s, about the same time as \verb+gcc -O+, which does not perform tail call optimization either.) We chose this core for benchmarking because it is very simple and performance or lack thereof is easy to explain.

On an x86-64 machine%
\footnote{Intel® Core™ i7-10610U CPU @ 1.80~GHz}, this procedure ran 10000 times on a 100000-element list takes 2.16~s using {\compcert}'s tail call elimination, and only 1.76~s (the same timing as \verb+gcc -O2+) if using tail recursion elimination, a 19\% reduction. This means that function prologue and epilogue can have nontrivial cost on small functions even on a highly out-of-order, speculative core. (Disabling tail call optimization results in 8.54~s.)

% CompCert list_traversal.c ecrins
% -fno-tailcalls -fno-tailrec 8.54
% -ftailcalls -fno-tailrec 2.16
% -ftailcalls -ftailrec 1.76
% gcc 9.17
% gcc -O 7
% gcc -O2 1.78

% gcc -O2
% .L12:
%        movq    %rdi, %rax
%        movq    8(%rdi), %rdi
%        testq   %rdi, %rdi
%        jne     .L12

%        ccomp -ftailcalls -ftailrec
%.L102:
%        movq    8(%rdi), %rax
%        cmpq    $0, %rax
%        je      .L103
%        movq    %rax, %rdi
%        jmp     .L102

Consider now our ``twisted quicksort'' procedure with the ordering $\prec_x$. We run it $m=10000$ times, with $i=0\dots m-1$, on an array of length 1000, initially containing the integers from 0 to 999999, for $x=1000000001-2i$, on the ``Rocket''.
Without tail call elimination, execution time is 37~s; it falls very slightly with tail call elimination, and drops to 34.4~s with tail recursion elimination.

We ran it with $m=100000$ times on the x86-64. Without tail recursion elimination, the timing is the same as with \verb+gcc -O+, but it increases with tail recursion elimination. \compcert's code generation is geared towards RISC processors with a reasonable number of registers (it does not exploit the memory-to-register operations of CISC processors) and it is possible that adding a loop path results in suboptimal register allocation. Moreover, the x86-64 is very complex and it is known that conditions such as the memory alignment of a loop body may have notable consequences on speed.

%  100 times, indexed by $i$, on an array of length 1000000
% x86-64
% gcc 8.83 9.17
% gcc -O 5.38
% gcc -O2 5.78
% ccomp
% notailrec 4.77
% tailrec 5.37
% notailcalls 4.91

% RISC-V
% #define n 1000
% #define m 100000

% gcc 32m30.921s
% gcc -O  11m29.632s
% gcc -O2 11m28.121s
% ccomp -fno-tailcalls 12m48.260s
% ccomp -ftailcalls -fno-tailrec 12m44.457s

% 10000 sorts of length 1000, Rocket
% ccomp -ftailcalls -ftailrec 34.421s
% ccomp -ftailcalls -fno-tailrec 36.803s
% ccomp -fno-tailcalls -fno-tailrec 36.979s
% gcc 1m27.541s
% gcc -O 0m29.805s
% gcc -O2 0m24.997s

% X86-64 ecrins
% gcc 1.94
% gcc -O 0.93
% gcc -O2 0.80
% ccomp -ftailcalls -fno-tailrec 0.92
% ccomp -ftailcalls -ftailrec 1.15
% ccomp -fno-tailcalls 0.92

\section{Conclusion, Related Work, and Perspectives}
Proving the correctness of program transformations involving a block-based memory model is often difficult. For several transformations available in mainstream compilers, but previously not available in formally verified compilers, we identified the key simulation arguments and successfully implemented these transformations into {\compcert}. We have successfully run hundreds of test cases using these transformations.

Some other improvements to {\compcert} \cite{DBLP:conf/icooolps/Gourdin23,DBLP:conf/cpp/SixGBMFN22,Gourdin_AFADL21,DBLP:journals/pacmpl/SixBM20,OOPSLA_2023,DBLP:phd/hal/Six21,DBLP:conf/popl/TristanL10,DBLP:phd/hal/Tristan09,DBLP:conf/popl/TristanL08}, starting with the register allocator \cite{DBLP:conf/cc/RideauL10}, instead of proving the total correctness of transformation passes, use \emph{translation validation}: a formally verified checker is run after the pass and checks that the source and target programs match. This checker may be helped by suitable annotations.

Translation validation has also been successfully applied, on select architectures to verify that the object code implementing the seL4 kernel matches the source code.
Similarly, Alive2 \cite{DBLP:conf/pldi/LopesLHLR21} checks that the result of a LLVM transformation matches the source.
In both cases, the checking is done by a SMT solver. In our context, that would entail either trusting the SMT solver, or getting from it some kind of certificate that could be used in a verified checker such as the one inside SMTCoq \cite{DBLP:conf/cav/EkiciMTKKRB17}, assuming we find one that supports a rich enough theory. This however seems a heavyweight solution, both because the solver needs to be run during every compilation run, and because of the interfacing code needed.

The formal semantics of LLVM were formalized as part of Vellvm, and illustrated with the formal verification of a pass for hardening programs against memory violations \cite{DBLP:conf/popl/ZhaoNMZ12}, a formally verified version of the SoftBound protection~\cite{DBLP:conf/pldi/NagarakatteZMZ09}.
Their system attaches to each pointer the bounds of the zone to which it can legally point to. Because affixing this metadata to the pointer itself would change the number of bytes needed to hold a pointer, leading to compatibility issues, they store the metadata in lookup tables in separate memory blocks.
One difference with our approach for canaries is that we change the layout of memory blocks used by the program, by extending stackframe blocks, whereas they keep the original memory blocks.
Memory metadata is kept in a disjoint memory space, the access to which is axiomatized.

One difficulty with pointer authentication is that, in the way that we axiomatized it, it does not respect memory injections. This is not an issue in our case, because we add pointer authentication late in the compilation process: the phase that deals with return addresses is not followed by phases whose correctness relies on injections.
It would be an issue if we wanted to allow arbitrary code and data pointers to be encoded through special qualifiers for pointers, as allowed by {\llvm} under MacOS~\cite{Apple_PAC_LLVM}.
In order to have pointer authentication respect memory injections, we could add a new kind of value to {\compcert}'s \verb+val+ type, for encoded authenticated pointers. This would however likely entail pervasively modifying {\compcert}, since this type is used pretty much everywhere; and thus we might end up with a fork of {\compcert} that would be hard to synchronize with the official releases. We may however study this approach in the future.

\citet{torrini:hal-03555551} proposed a more ambitious view of a {\compcert} backend with a semantics incorporating symbolic encryption, for compilation targeting a special processor with built-in code encryption.

The abstract vision of pointers as pairs $(b,o)$ of block identifier and offset has advantages, but also limitations. For once, the semantics of programs doing bit manipulations on pointers (such as storing data in low-order bits of pointers known to be multiples of 2, 4 or 8) is undefined. It also unnaturally views the stack as a list of independent blocks, whereas in reality they are contiguous in memory, which has some adverse consequences for efficiency ({\compcert} stores a pointer to the next stack frame in the current stack frame).
Some alternative models viewing pointers as integers were developed for {\compcert}~\cite{DBLP:conf/itp/BessonBW15,DBLP:journals/jar/BessonBW19a}.

CHERI \cite{DBLP:conf/isca/WoodruffWCMADLNNR14} proposes an execution model where pointers can carry \emph{capabilities}, including buffer bounds, that are enforced in hardware. With such a system, one can render buffer overflows impossible. An extra bit, unforgeable in software, distingues capabilities from normal data. CHERI has been successfully implemented as the ``Morello'' variant of the AArch64 architecture \cite{DBLP:journals/micro/GrisenthwaiteBWMSW23}. At minimum, a compiler for a CHERI-enhanced platform must cope with 32-bit pointers being replaced by 65-bit pointers, 64-bit pointers by 129-bit pointers, and provide memory copy operations that copy the capability bits.
For CHERI to be useful against buffer overflows, pointers to memory objects must be created with capabilities limited to these objects. 
Finer-grained security could also involve further restricting pointer capabilities, especially block sizes, automatically or through explicit program specification (for instance, one may want to restrict a function being called to operate only on part of an object, such as an array slice). In any case, supporting CHERI in {\compcert} would entail substantial and pervasive changes.

Some high-security or high-safety systems duplicate information and/or computations: mismatches indicate hardware faults, memory corruption, or malicious attack. Duplicating computations can be done in hardware (run two copies of the same core in lockstep and check if their output pins match, trap if they don't) or in software, either by modifying the source code or through a compilation pass \cite{DBLP:journals/tr/OhSM02,DBLP:conf/cgo/ReisCVRA05} (as in STMicroelectronics's modified LLVM, \emph{SecSwift} \cite{Ferriere_EUROLLVM2019}). We envision such a duplicating pass could be added to {\compcert}. Duplicating information in memory at compilation time is however more difficult, at least if compiling from C or a language similar to it, because a pointer designates one single block of memory, not one block and its copy.

Our proofs of correctness of phases that insert security countermeasures just show that they preserve legal executions. We do not prove that these countermeasures have any effect on any attack. Doing so would probably involve incorporating a threat model inside the semantics of the program and show properties such as ``if a program executes a branch in the attack semantics that is not executed in the original semantics, then its execution must stop before issuing an observable event''. While pen-and-paper proofs of such \emph{adequation} properties on toy languages are feasible, formal proofs for them in the context of semantics of compiler intermediate representations are a challenging task.

In the case of pointer authentication, one further difficulty is that this proof of security would have to be relative to the security of the encoding and decoding primitives, in much the same way that the security of a cryptographic protocol is generally not proved in absolute terms, but rather relative to the security of the primitives involved, often through some \emph{game reductions} (if the attacker can win against the protocol, then, up to some margin in probabilities, it can win against the primitives).

One goal of our work was to minimally change {\compcert}, so that our additions could be presented as independent passes that could eventually be submitted for integration into the official releases. This was an important constraint. Many of the extensions or alternate solutions described in this section entail modifying global definitions (such as the value type) with possibly far-reaching consequences.

%% Local Variables:
%% TeX-master: "memory_simulation_ACM.tex"
%% End:

\section*{Acknowledgments}
This work is partially supported by the \href{https://www.pepr-cyber-arsene.fr/}{ARSENE} project funded by the “France 2030” government investment plan managed by the \grantsponsor{ANR}{French National Research Agency (ANR)}{https://anr.fr/}, under the reference \grantnum[https://www.pepr-cybersecurite.fr/projet/arsene/]{ANR}{ANR-22-PECY-0004}.

\printbibliography
\end{document}